\documentclass[conference]{IEEEtran}

\usepackage{cite}
\usepackage{amsmath,amssymb,amsfonts}
\usepackage{algorithmic}
\usepackage{graphicx}
\usepackage{textcomp}
\usepackage{xcolor}
\usepackage[hidelinks]{hyperref}
\usepackage{booktabs}
\usepackage{harveyballs}
\usepackage{tikz}
\usetikzlibrary{positioning, arrows.meta, calc, shapes.geometric}
\usepackage{graphicx}
\usepackage[T1]{fontenc}
\usepackage{comment}
\usepackage[final,commandnameprefix=ifneeded]{changes}

\usepackage{todonotes}

\def\BibTeX{{\rm B\kern-.05em{\sc i\kern-.025em b}\kern-.08em
    T\kern-.1667em\lower.7ex\hbox{E}\kern-.125emX}}
\begin{document}

\title{Decentralized Multi-Stage Service Negotiation in Value Networks with Proactive AAS using Standardized Submodels}

\author{\IEEEauthorblockN{David Dietrich, Carsten Ellwein, Armin Lechler, Alexander Verl}
\IEEEauthorblockA{\textit{Institute for Control Engineering of Machine Tools and Manufacturing Units (ISW)}\\
\textit{University of Stuttgart}, Stuttgart, Germany \\
\{david.dietrich, carsten.ellwein, armin.lechler, alexander.verl\}@isw.uni-stuttgart.de}}

\maketitle

\begin{abstract}
Modern value networks require decentralized and scalable coordination of cross-company, service-based value chains, which are insufficiently supported by today’s predominantly centralized and custom-integrated solutions.
This paper addresses how realistic negotiation behaviors and multi-stage service interactions can be systematically integrated into proactive Asset Administration Shells (AAS) for cross-company planning based on standardized capability models.
Existing work does not yet provide a methodologically grounded active AAS architecture that combines the VDI/VDE 2193 Industry 4.0 Language, the emerging IDTA 02020 “Capabilities for Industrial Appliances” submodel, and explicit, configurable negotiation behavior in a scalable way.
First, we contribute a MaSE-based methodological design for the active AAS that refines roles into layered agent classes (communication, behavior, context), enabling reusable requester, provider, and proxy patterns.
Second, we realize a standards-compliant capability modeling and constraint framework that encodes multi-stage dependencies, public/private properties, and feasibility checks directly in the AAS submodel, and supports provider- and requester-side optimization strategies.
Third, we exemplify the approach in an industrial Manufacturing-X testbed with multi-stage product services and multiple companies, showing that different negotiation behaviors and proxy-based zoning lead to measurable differences in completion times and system scalability.

\end{abstract}


\section{Introduction} \label{sec:intro}

The increasing complexity and dynamics of modern value chains require new approaches to planning~\cite{10.1145/3232174.3232176} and coordination \replaced{across}{between} company boundaries~\cite{Dietrich.2024e}.
Central coordination is typically driven by the manufacturers of the original equipment and is supported by manual configuration or predefined systems \cite{Stadtler.2005}\replaced{, which}{ that} lead to limited scalability and high coordination effort when faced with increasing volatility~\cite{Lucker.2025}.
To address these limitations, concepts such as virtual enterprises \cite{CamarinhaMatos.1999} and cloud manufacturing~\cite{8436332} aim to improve flexibility through service-oriented architectures and shared digital infrastructures.
\deleted{According to the understanding of~\cite{Diedrich.06.10.2022}, a service encapsulates the offered or requested capabilities extended with commercial aspects.}
However, these approaches are largely based on centralized platforms that are limited by data or planning governance and require custom integration~\cite{Gharibvand.2024}.

To realize scalable, dynamic, 
and decentralized coordination of service-based value chains, the interaction \replaced{between}{of} companies requires interoperable standards.
The Asset Administration Shell (AAS) specifies such a standard for the digital representation of industrial assets. 
\replaced{It}{The AAS} specifies a metamodel, together with a defined API and use-case-oriented submodels that hold the relevant data. 
A decentralized interaction concept to realize a so-called proactive AAS is introduced in VDI/VDE 2193 with the Industry 4.0 Language (I4.0L) in~\cite{VereinDeutscherIngenieure.}.
The I4.0L proposes a vocabulary based on \added{the }AAS and semantic dictionaries\replaced{, defines}{ and} the structure of messages in~\cite{VereinDeutscherIngenieure.}, \replaced{and specifies additional}{as well as specific} \replaced{interaction}{bidding} protocols in further parts.
Currently, the only specified interaction protocol is the bidding protocol, which is based on the contract net protocol (CNP) and enables \deleted{the }decentralized \added{service }negotiation\deleted{ of services}~\cite{VereinDeutscherIngenieure.c}.
\replaced{In the specifications, no}{A} technological realization of the I4.0L or a semantic definition of the bidding procedure is \deleted{not }provided\deleted{ in the specifications}.
To realize a proactive AAS, current architectural considerations extend the passive AAS by\added{ adding} an active part containing a component manager, algorithms, an interaction manager, and a messenger~\cite{Belyaev.2019}.
A current enabler \replaced{for}{in} the standardized negotiation of multi-stage services exists in the newly released \deleted{submodel specification, }``Capability Description''\added{ submodel specification}~\cite{idta02020capabilities}.
\deleted{The submodel structures the capabilities of an asset into sets, each described by a container with a capability identifier, properties, relations, and a human-readable description.
The relations may contain constraints, compositions, generalizations, and realizations.}

\added{As terminology partially varies in the state of the art, the following definitions based on the information model for capabilities, skills, and services in \cite{Diedrich.06.10.2022} are used.
In that understanding, \textit{Capabilities} abstract production functions.
The implementation of capabilities is realized with \textit{Skills} on a specific resource, that provide an interface that can \textit{execute} the skills with given parameters.
The offering or request of capabilities in a supply chain is abstracted by \textit{Services} that encapsulates them with commercial aspects.
Capabilities as well as services can be \textit{offered} or \textit{requested}.
}

Although the adoption of\added{ the} AAS as\added{ a} digital representation is moving forward, only a few works address or implement proactive AAS according to existing standards, as the following state of the art in \autoref{sec:sota} shows.
Furthermore, our preliminary work in \cite{Dietrich.2026} shows limitations\added{that arise} at scale \replaced{when}{assumed in the absence of }negotiation behaviors\added{ are absent}.
This raises the research question: 
\replaced{How can proactive AAS achieve decentralized planning of cross-company value chains based on services? With a focus on the following subquestions: (i) How can different negotiation behaviors be integrated into proactive AAS architectures? (ii) How can a standardized exchange of realistic service interactions be built upon the standardized submodel for capability description?}{How can different negotiation behaviors and realistic service interactions for the decentral planning of cross company value chains based on the capabilities submodel be integrated in proactive AAS.}
To answer this research question, \autoref{sec:sota} further details the state of the art.
Oriented towards the methodology of \replaced{M}{m}ultiagent \replaced{S}{s}ystems \replaced{E}{e}ngineering (MaSE)\added{presented} in \cite{DELOACH.2001}, the following sections extend the current design of proactive AAS in \autoref{sec:meth}, describe the realization in \autoref{sec:real}, and exemplify \added{the approach through }a use case based on a function demonstrator in \autoref{sec:exe}.
The results are 
concluded in \autoref{sec:conc}.

\section{State of the Art} \label{sec:sota}

Various research has been conducted in the field of proactive AAS \added{(also referred to as Type~3 AAS)}\replaced{, whilst}{ and} interpretations \added{and definitions }of the term vary widely~\cite{Ellwein.2026}. 
For the realization of service negotiation via proactive AAS, \cite{Stolze.2024b} introduces architectural approaches and technological interfaces.
A monolithic, agent-based, and orchestrating architecture for the integration of the active AAS part \replaced{is}{are} discussed, and communication mechanisms via MQTT and REST with regard to events and operations of the AAS metamodel and its API are evaluated.
Automation of production planning using business process modeling notation (BPMN) and custom AAS submodels is introduced in \cite{Stolze.2024}.
It reflects the negotiation of a single service and uses AAS operations to receive messages according to the I4.0L.
Similarly, \cite{Grunau.2022b} introduces a bidding app based on custom AAS and MQTT to realize an order\replaced{-}{ }driven production based on the I4.0L.
It furthermore evaluates the integration of the active AAS part in the AAS-Server compared to a separate application.
An integration of proactive AAS and multi-agent systems (MAS) within the reference architectural model industry 4.0 is presented in \cite{Sakurada.2023} using the Java Agent Development framework and CNP for interaction.
To achieve \deleted{a }decentralized coordination, a behavior for multi-stage products and resources \replaced{is}{are} integrated into the interaction.
Another MAS based on AAS for dynamic replanning is presented in \cite{Jungbluth.2022}, which enables product and resource agents to manage unforeseen events based on capabilities and skills.
The matching of products and resources is realized by a feasibility check to identify alternative, available resources.
An extension of this approach to a cross-enterprise shared production scenario is introduced in \cite{Jungbluth.2023}, utilizing I4.0L and Gaia-X to facilitate automated bidding procedures.
This framework leverages custom AAS submodels, such as Part-Specification and Assured Services, to extract product features and map them to manufacturer services within a sovereign and trusted data space.
A different proactive AAS interaction protocol for ``Verification of Request'', according to NAMUR NE 178 in the process industry, is implemented and evaluated in \cite{Pakala.2025}, showing \deleted{the }controlled and semantically interoperable negotiation\added{s} for secure write access in automation systems.
As examples are usually conducted with a limited number of assets, the modeling of a proactive AAS system with negotiation according to I4.0L and scalability criteria is presented in \cite{Dietrich.2026}.
The analysis performed on scale shows limitations that require further attention to realize a scalable behavior.
%
%
The challenges identified in the introduction, together with the analysis of the state of the art, form the basis for deriving a set of requirements that serve as evaluation criteria.
As foundations, the service abstraction, the proactive architecture, the I4.0L compliance, and a standardized submodel (e.g., \cite{idta02020capabilities}) are set.
Further requirements for the cross\replaced{-}{ }company negotiation of multi-stage services are considered as the multi-stage planning itself, the detailed regard of negotiation behavior and algorithms, an initial discovery of companies or services, and an evaluation of the scalability.
A comparison of the state of the art with regard to evaluation criteria is shown in \autoref{tab:sota}. 

\begin{table}[t]
    \centering
    \caption{Evaluation of the state of the art}
    \resizebox{\columnwidth}{!}{
    \begin{tabular}{l c c c c c c c c}
         Requirements & \cite{Stolze.2024b} & \cite{Stolze.2024} & \cite{Grunau.2022b} & \cite{Sakurada.2023} & \cite{Jungbluth.2022} & \cite{Jungbluth.2023} & \cite{Pakala.2025} & \cite{Dietrich.2026}\\
         \midrule
         Service abstraction \cite{Diedrich.06.10.2022} & \harveyBallNone & \harveyBallHalf & \harveyBallHalf & \harveyBallHalf & \harveyBallNone & \harveyBallFull & \harveyBallNone & \harveyBallFull \\
         Proactive architecture \cite{Belyaev.2019} & \harveyBallFull & \harveyBallFull & \harveyBallNone & \harveyBallNone & \harveyBallNone & \harveyBallNone & \harveyBallNone & \harveyBallNone \\
         I4.0L compliant \cite{VereinDeutscherIngenieure.} & \harveyBallNone & \harveyBallFull & \harveyBallFull & \harveyBallHalf & \harveyBallFull & \harveyBallHalf & \harveyBallHalf & \harveyBallFull \\
         Standardized Submodel & \harveyBallNone & \harveyBallNone & \harveyBallNone & \harveyBallNone & \harveyBallNone & \harveyBallNone & \harveyBallNone & \harveyBallNone \\
         Multi-stage planning & \harveyBallNone & \harveyBallNone & \harveyBallNone & \harveyBallFull & \harveyBallHalf & \harveyBallHalf & \harveyBallNone & \harveyBallNone \\
         Detailed behavior \& algorithms & \harveyBallNone & \harveyBallHalf & \harveyBallHalf & \harveyBallHalf & \harveyBallHalf & \harveyBallHalf & \harveyBallHalf & \harveyBallHalf \\
         Initial discovery & \harveyBallNone & \harveyBallHalf & \harveyBallFull & \harveyBallFull & \harveyBallFull & \harveyBallFull & \harveyBallNone & \harveyBallFull\\
         Scalability evaluation & \harveyBallNone & \harveyBallNone & \harveyBallNone & \harveyBallHalf & \harveyBallNone & \harveyBallHalf & \harveyBallNone & \harveyBallHalf
    \end{tabular}}
    \label{tab:sota}
    \deleted{\vspace{-15pt}}
\end{table}

Even though no standardized capability submodel was available so far, it is remarkable that none of the approaches intends to use existing ones for, e.g., technical data.
Furthermore, none of the approaches fulfills the other foundations.
With regard to the further requirements, only\added{ a} few works address multi-stage planning or an evaluation of scalability.
Whilst an initial discovery is mostly existing due to MQTT with a publish\replaced{-}{ }subscribe principle, most works do not present a detailed behavior or the algorithms of the service provider or requester. 
This underscores the need for scalable approaches for service interaction based on the AAS.



\section{Analysis and Design of the proactive AAS} \label{sec:meth} 

Based on the state of the art, the analysis and design of the proactive AAS are presented in this chapter.
The methodology in the following subsections is aligned with the multi-agent systems engineering approach in \cite{DELOACH.2001}, starting with the capture of goals as the first step of the analysis.
%

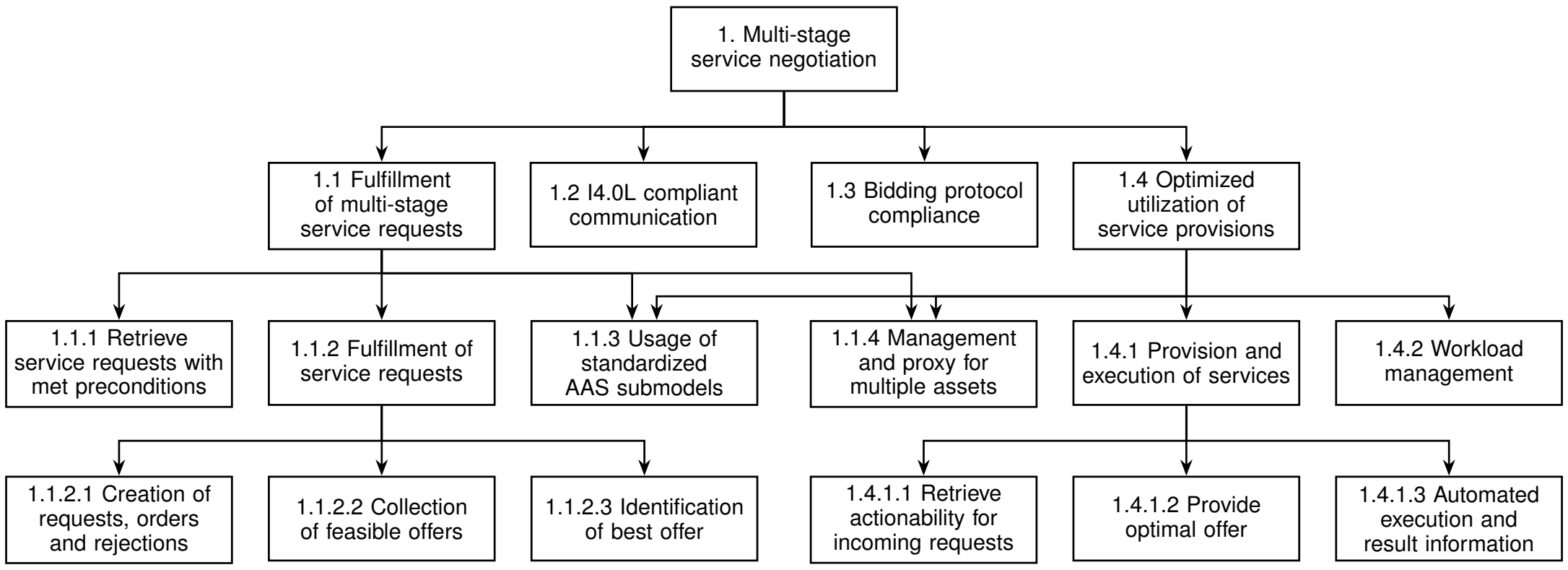
\begin{figure*}[t]
    \centering
    \resizebox{0.8\textwidth}{!}{
    \begin{tikzpicture}[
    node distance=1cm and 0.5cm,
    block/.style={
        rectangle, 
        draw=black, 
        thick, 
        fill=white,
        text width=3cm, 
        align=center, 
        minimum height=1.2cm,
        font=\sffamily\small
    },
    root/.style={
        block,
    },
    line/.style={
        draw, 
        -Stealth, 
        thick
    }
]

\node [root] (n1) {1. Multi-stage service negotiation};

\node [block, below left=1cm and 2.5cm of n1] (n11) {1.1 Fulfillment of multi-stage service requests};
\node [block, below left=1cm and -1.25cm of n1] (n13) {1.2 I4.0L compliant communication};
\node [block, below right=1cm and -1.25cm of n1] (n14) {1.3 Bidding protocol compliance};
\node [block, below right=1cm and 2.5cm of n1] (n12) {1.4 Optimized utilization of service provisions};

\node [block, below left=1cm and 0.5cm of n11] (n111) {1.1.1 Retrieve service requests with met preconditions};
\node [block, below=1cm of n11] (n112) {1.1.2 Fulfillment of service requests};
\node [block, below right=1cm and 0.5cm of n11] (n113) {1.1.3 Usage of standardized AAS submodels};
\node [block, below left=1cm and 0.5cm of n12] (n114) {1.1.4 Management and proxy for multiple assets};

\node [block, below left=1cm and 0.5cm of n112] (n1121) {1.1.2.1 Creation of requests, orders and rejections};
\node [block, below=1cm of n112] (n1122) {1.1.2.2 Collection of feasible offers};
\node [block, below right=1cm and 0.5cm of n112] (n1123) {1.1.2.3 Identification of best offer};

\node [block, below=1cm of n12] (n121) {1.4.1 Provision and execution of services};
\node [block, below right=1cm and 0.5cm of n12] (n122) {1.4.2 Workload management};

\node [block, below left= 1cm and 0.5cm of n121] (n1211) {1.4.1.1 Retrieve actionability for incoming requests};
\node [block, below= 1cm of n121] (n1212) {1.4.1.2 Provide optimal offer};
\node [block, below right= 1cm and 0.5cm of n121] (n1213) {1.4.1.3 Automated execution and result information};

\path [line] (n1.south) -- ++(0,-0.5) -| (n11.north);
\path [line] (n1.south) -- ++(0,-0.5) -| (n12.north);
\path [line] (n1.south) -- ++(0,-0.5) -| (n13.north);
\path [line] (n1.south) -- ++(0,-0.5) -| (n14.north);

\path [line] (n11.south) -- ++(0,-0.33) -| (n111.north);
\path [line] (n11.south) -- (n112.north);
\path [line] (n11.south) -- ++(0,-0.33) -| ([xshift=-5pt]n113.north);
\path [line] (n11.south) -- ++(0,-0.33) -| ([xshift=-5pt]n114.north);

\path [line] (n112.south) -- ++(0,-0.5) -| (n1121.north);
\path [line] (n112.south) -- (n1122.north);
\path [line] (n112.south) -- ++(0,-0.5) -| (n1123.north);

\path [line] (n12.south) -- ++(0,-0.66) -| ([xshift=5pt]n113.north);
\path [line] (n12.south) -- ++(0,-0.66) -| ([xshift=5pt]n114.north);
\path [line] (n12.south) -- (n121.north);
\path [line] (n12.south) -- ++(0,-0.66) -| (n122.north);

\path [line] (n121.south) -- ++(0,-0.5) -| (n1211.north);
\path [line] (n121.south) -- (n1212.north);
\path [line] (n121.south) -- ++(0,-0.5) -| (n1213.north);

\end{tikzpicture}
    }
    \caption{System hierarchy}
    \label{fig:goal}
    \deleted{\vspace{-10pt}}
\end{figure*}

\paragraph{Capturing Goals}
\replaced{Based on the context of multi-stage service negotiation introduced in value networks, the system objectives were derived by decomposing this negotiation problem into goals that address distinct concerns while remaining mutually dependent.
These are structured in the goal hierarchy shown in \autoref{fig:goal}. The main goal is split into a requester side goal (1.1) and a provider side goal (1.4), since both parties pursue different objectives during the negotiation.
Both branches are constrained by compliant communication (1.2) and bidding protocol compliance (1.3), which apply equally to requester and provider and therefore sit above both branches in the hierarchy.\\
On the requester side, a request cannot be evaluated as a single decision, because it spans multiple stages whose preconditions are only satisfied progressively.
For this reason, checking these preconditions (1.1.1) is treated as a separate goal from the actual fulfillment.
Fulfillment (1.1.2) depends on two different kinds of information that need to be established one after another.
First, what can legitimately be requested, covering requests, orders, and rejections. Second, what is actually offered in return, which requires collecting feasible offers and then identifying the best one among them.\\
The provider side (1.4) follows a comparable logic. Since not every incoming request can be serviced, the provider must first determine whether a request is actionable (1.4.1.1) before formulating an offer (1.4.1.2) and, once it is accepted, executing it (1.4.1.3).
Workload management (1.4.2) is introduced as its own goal because optimal utilization cannot be judged by looking at single requests in isolation.
It only becomes meaningful when requests are considered together against the provider's overall capacity.\\
Both sides also rely on two shared foundations.
To build on existing specifications, standardized AAS submodels (1.1.3) need to form the basis of the proactive part, even if they are not mandatory for a compliant communication on the top level.
To enable the grouping of assets in hierarchical system, a coordination across multiple assets (1.1.4) is required.
The fulfillment of single services is the point where requester and provider decisions converge, requiring dedicated algorithms to reconcile both perspectives.}{
Based on the context of multi-stage service negotiation introduced in value networks, the system objectives were identified and are structured in the goal hierarchy in \autoref{fig:goal}.
With the multi-stage service negotiation as main goal, different branches for their request (1.1) and provision (1.3) are identified, connected through a compliant communication (1.2).
While the multi-stage dependency (1.1.1) shall be resolved for the fulfillment of single requests (1.1.2) for the requester, the provision of single services (1.3.1) within a production system capacity (1.3.2) is of relevance to the provider.
Both are based on standardized AAS (1.1.3) and need coordination across multiple assets (1.4).
For the fulfillment of single services, various algorithms are required.}

\paragraph{Applying Use Cases}
\replaced{To validate the goal hierarchy, a representative use case was derived from the system context and translated into sequence diagrams. The use case considers a requested service with multiple stages, performed within a decentralized network of service providers, and illustrates how the required multi-stage services, together with workload information and commercial aspects, can be described using the submodel "Capabilities for Industrial Appliances" \cite{idta02020capabilities}. In line with the scalability and decentralization goals identified above, the use case distinguishes between a requester and a provider perspective, which are illustrated separately in the following.}{In the second step, the use cases are derived from the system context and transferred into sequence diagrams.
As an abstract use case, a requested service with multiple stages shall be performed in a decentral network of service providers.
The required multi-stage services and the services provided with possible workload information and commercial aspects shall be described with the submodel ``Capabilities for Industrial Appliances``~\cite{idta02020capabilities}.
To achieve a scalable and decentral system, a differentiation in the requester and in the provider perspective is made.}

\begin{figure}[t]
    \centering
    \includegraphics[width=\linewidth]{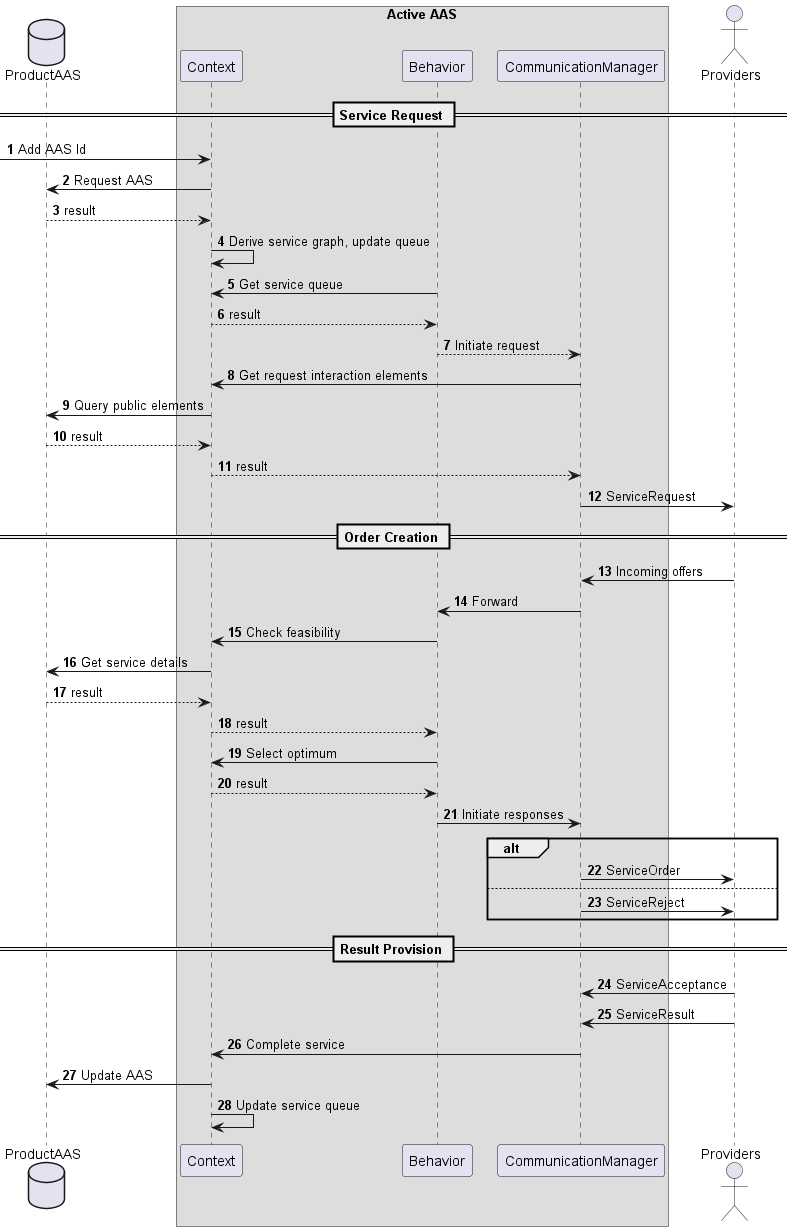}
    \caption{Requester sequence diagram}
    \label{fig:req}
    \deleted{\vspace{-15pt}}
\end{figure}

\replaced{The sequence diagram in \autoref{fig:req} illustrates the requester perspective, focusing on the successful flow of the process. Based on the AAS-Id of a passive AAS, required services and their dependencies are derived from their context, and the validity of these dependencies is checked (\autoref{fig:req} Nos. 1-4). Services whose preconditions are met are then negotiated based on the behavior of the active AAS part (\autoref{fig:req} Nos. 4-7), and public attributes of the service are retrieved to create an I4.0L compliant service request (\autoref{fig:req} Nos. 8-12). For incoming offers, an initial check verifies that they are formally compliant and submitted within the required response time (\autoref{fig:req} Nos. 13-14), before feasibility is assessed using the approach introduced in \cite{Dietrich.2026}, based on AAS properties and constraints (\autoref{fig:req} Nos. 15-18). Based on target criteria, an optimal offer is selected and corresponding orders and rejections are issued (\autoref{fig:req} Nos. 19-23). If no feasible offers are found, a negotiation retry may be triggered depending on the behavior configuration. Otherwise, order acceptance and the service result are awaited and updated in the AAS (\autoref{fig:req} Nos. 24-27). Finally, dependencies of pending service requests are refreshed (\autoref{fig:req} No. 28), closing the loop of multi-stage planning back to No. 5, until all requests of an AAS are fulfilled or identified as unachievable, for example due to retry limits.}{The sequence diagram of the requester is shown in \autoref{fig:req} and is explained below, focusing on the successful flow of the process.
Based on the AAS-Id of a passive AAS, required services and their dependencies are derived based on their context and the validity of dependencies is validated (\autoref{fig:req} Nos. 1-4). 
Services with preconditions met are then negotiated based on the behavior of the active AAS part (\autoref{fig:req} Nos. 4-7).
Public attributes of the service are retrieved to create an I4.0L compliant service request (\autoref{fig:req} Nos. 8-12).
For incoming offers, an initial check is conducted to ensure that they are formally compliant and submitted within the required response time (\autoref{fig:req} Nos. 13-14).
The feasibility check is then carried out, as introduced in \cite{Dietrich.2026} based on the AAS properties and constraints (\autoref{fig:req} Nos. 15-18).
Based on target criteria, an optimal offer is selected and orders and rejects are initiated as responses (\autoref{fig:req} Nos. 19-23).
In case no feasible offers are found, a negotiation retry may be initiated based on the behavior configuration.
Otherwise, the order acceptance and the service result are awaited and updated in the AAS (\autoref{fig:req} Nos. 24-27).
Finally, dependencies of pending service requests are refreshed (\autoref{fig:req} No. 28) closing the loop of multi-stage planning to No. 5 until all service requests of an AAS are fulfilled or unachievable services due to, e.g., retry limits are met.}

\begin{figure}[t]
    \centering
    \includegraphics[width=\linewidth]{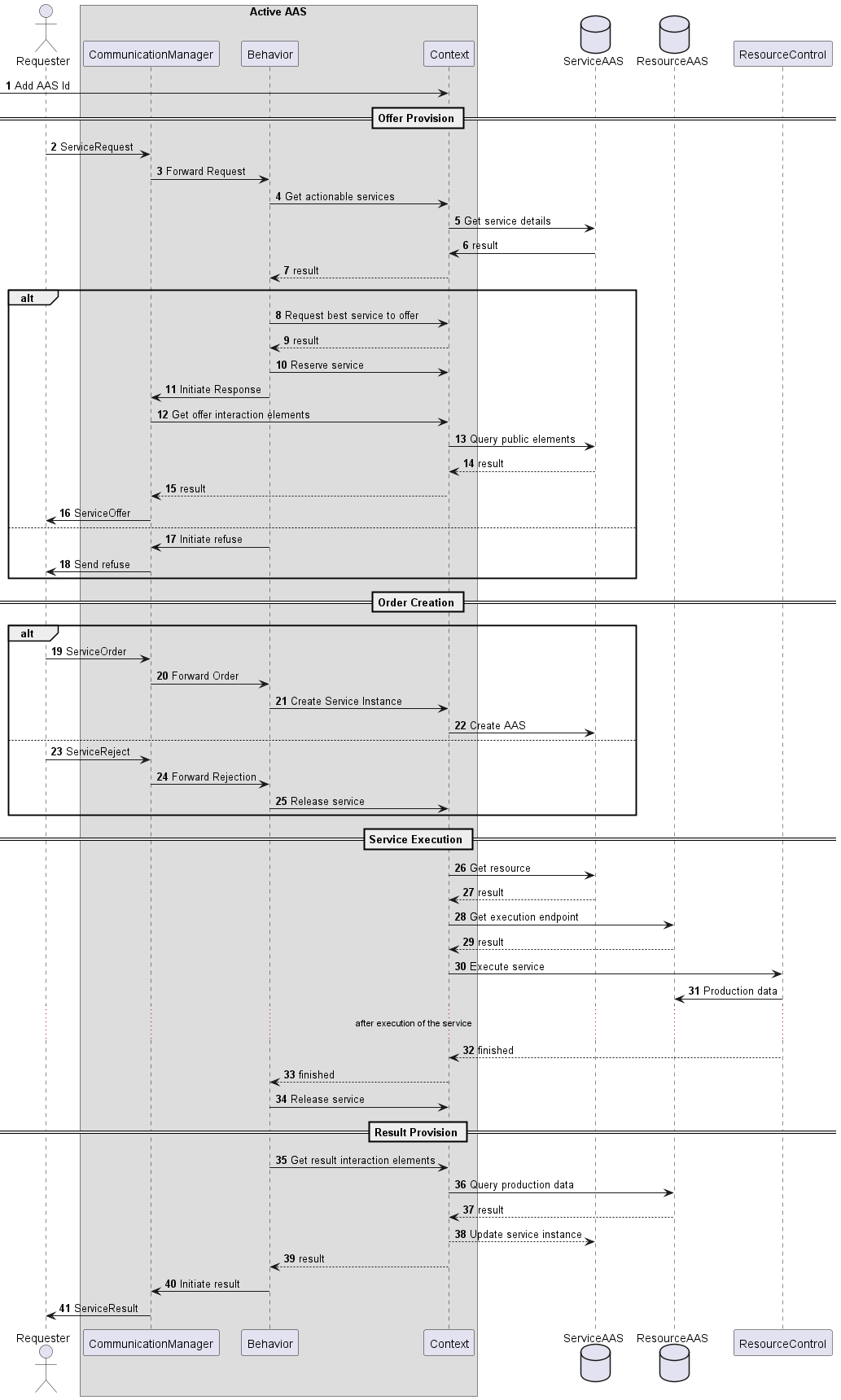}
    \caption{Provider sequence diagram}
    \label{fig:prov}
    \deleted{\vspace{-15pt}}
\end{figure}

\replaced{The provider perspective is illustrated analogously in \autoref{fig:prov}. Following the registration of a passive type AAS (\autoref{fig:prov} No. 1), the capability and actionability of the offered services are assessed (cf. \cite{Dietrich.2026}) (\autoref{fig:prov} Nos. 2-7). Depending on the behavior configuration, the best service to offer is identified, particularly when multiple requests arrive within a response timeout (\autoref{fig:prov} Nos. 8-9). The chosen service is reserved in line with the binding of offers described in \cite{VereinDeutscherIngenieure.c}, and a response is issued based on the AAS's public properties (\autoref{fig:prov} Nos. 10-16). Capable but non-actionable services are refused (\autoref{fig:prov} Nos. 17-18). Once an offer is accepted within the response time, a service instance is created (\autoref{fig:prov} Nos. 19-22). If offers are rejected or no response is received, the service is released again for other requests (\autoref{fig:prov} Nos. 23-25). To execute the service, a referenced resource skill is retrieved and run on the resource controller (\autoref{fig:prov} Nos. 26-30), during which production data is added to the resource AAS (\autoref{fig:prov} No. 31), until execution is complete and the service is released (\autoref{fig:prov} Nos. 32-34). The service results are then aggregated from the resource AAS into the service instance and provided to the requester via an I4.0L message (\autoref{fig:prov} Nos. 35-41).}{The provider perspective of the service negotiation is shown in \autoref{fig:prov}.
After initial registration of passive type AAS (\autoref{fig:prov} No. 1), the capability and actionability (cf. \cite{Dietrich.2026}) of the services are processed (\autoref{fig:prov} Nos. 2-7).
Depending on the behavior configuration, the best service to offer is identified, especially if multiple requests are incoming within a response timeout (\autoref{fig:prov} Nos. 8-9).
The chosen service is then reserved due to the binding of offers in \cite{VereinDeutscherIngenieure.c} and responses are initiated based on the public properties of the AAS (\autoref{fig:prov} Nos. 10-16).
In case of capable, but not actionable services, a refuse is sent (\autoref{fig:prov} Nos. 17-18).
As soon as an offer is accepted within the response time, a service instance is created (\autoref{fig:prov} No. 19-22).
If offers are rejected or no response is received, the service is released again to be actionable for other requests (\autoref{fig:prov} Nos. 23-25).
To run the service, a referenced resource skill from the service is retrieved and executed on the resource controller (\autoref{fig:prov} Nos. 26-30).
During the process production data are added to the resource AAS (\autoref{fig:prov} No. 31), until the execution is finished and the service is released (\autoref{fig:prov} No. 32-34).
To provide the results of the service to the requester, the interaction elements of the I4.0L message are aggregated from the resource AAS, updated in the service instance and provided via an I4.0L message (\autoref{fig:prov} Nos. 35-41).}

\added{Together, both perspectives demonstrate how the goals identified in the last section are realized in a concrete negotiation process, and confirm that the derived use case satisfies the requester and provider side objectives.}

\paragraph{Refining Roles}
\replaced{Following up\deleted{ on the captured goals and the sequence diagrams}, roles, and concurrent tasks are refined in the third and last step of the analysis phase, as this step is what translates the abstract goals of a) into concrete responsibilities that can later be assigned to software components. The resulting roles, their description, task, and reference to the goals of a) are displayed in \autoref{tab:roles}, so that each role can be traced back to the specific sub-goal it fulfills rather than being introduced independently. As the objectives are not further detailed for clarity, some roles span multiple objectives, but are categorized by their primary function in the table.
With regard to the roles, their coherence with the proactive AAS architecture in \cite{Belyaev.2019} is what justifies structuring the subsequent agent classes around them. The messenger and interaction manager are equally represented, since both concern message exchange independently of the negotiation logic. The AAS-handler creates the connection to the passive part, which is required for any role to access asset specific information at all. The provision manager, precondition check, and negotiator mainly manage the component, since they implement the decision logic identified in the goal hierarchy. The constraint check, optimizer, and execution manager are mainly algorithms interfacing with the latter, as they provide the computational means by which those decisions are actually made. The proxy role, referring to agent systems including zones or supernodes, introduces a new component that is not part of the existing architecture in \cite{Belyaev.2019}, and is required to support the coordination across multiple assets identified as goal 1.1.4. This alignment between roles and existing architecture is what allows the role table to serve as the direct basis for the derivation of the agent classes, rather than as a separate design artifact.}{Following up on the captured goals, the sequence diagrams, roles, and concurrent tasks are refined in the third and last step of the analysis phase.
The resulting roles, their description, task, and reference to the goals of a) are displayed in \autoref{tab:roles}.
As the objectives are not further detailed for clarity, some span multiple roles but are categorized by their primary function in the table.\\
With regard to the roles, their coherence with the proactive AAS architecture in \cite{Belyaev.2019} is visible: The messenger and interaction manager are equally represented; the AAS-handler creates the connection to the passive part; the provision manager, precondition check, and negotiator mainly manage the component; the constraint check, optimizer, and execution manager are mainly algorithms inferring with the latter.
The proxy role referring to agent systems, including zones or supernodes, introduces a new component.}

\begin{table}[t]
    \centering
    \caption{Role overview}
    \resizebox{\linewidth}{!}{
    \begin{tabular}{p{0.3cm} p{1.3cm} p{2.5cm} p{2.5cm} p{0.8cm}}
        Nr. & Role & Description & Tasks & Goal\\
        \midrule
        R1 & Messenger & I4.0L compliant message exchange, e.g., via REST, MQTT or XMPP & Send and receiving valid I4.0L messages & 1.2 \\
        R2 & Interaction manager & Ensuring bidding-protocol compliant messages & Tracking of conversations, assurance of protocol state machine and response times & 1.3 \\
        R3 & AAS-handler & Extraction of service information from the AAS & Processing of properties and references of requested and provided services & 1.1.3\\
        R4 & Proxy & Coordination of multiple assets & Proxy of messages, management and optimization & 1.1.4 \\
        R5 & Provision manager & Management of workload and service reservations & Workload check, reservation \& release & 1.4.2\\
        R6 & Precondition check & Identify services possible to negotiate & Validate transition constraints, update preconditions met & 1.1.1 \\
        R7 & Negotiator & Negotiation of service according to desired business behavior & Order processing, retry mechanism & 1.1.2, 1.4.1, 1.1.2.1 \\
        R8 & Constraint check & Identification of capability, actionability and feasibility & Verification of property constraints & 1.1.2.2, 1.4.1.1 \\
        R9 & Optimizer & Derive best match of $m$ requests to $n$ provisions & Sorting based on target criteria & 1.1.2.3, 1.4.1.2 \\
        R10 & Execution Manager & Execution of negotiated services & Resource derivation, skill handling, e.g. via OPC UA & 1.4.1.3      
    \end{tabular}}
    \label{tab:roles}
    \deleted{\vspace{-15pt}}
\end{table}

\begin{figure}
    \centering
    \resizebox{\linewidth}{!}{

\begin{tikzpicture}[
    layer_box/.style={draw, thick, minimum width=16cm, minimum height=2cm, align=left, anchor=north west},
    module/.style={draw, rectangle, rounded corners=2pt, minimum width=2.2cm, minimum height=1cm, align=center, fill=white},
    db_shape/.style={cylinder, draw, shape border rotate=90, aspect=0.25, minimum width=2.5cm, minimum height=1.5cm, fill=white, thick},
    interface/.style={circle, draw, thick, minimum size=4mm, fill=white},
    label_style/.style={font=\large\bfseries, align=right},
    arrow/.style={-Stealth, thick},
    mydash/.style={-Stealth, dashed, draw=gray}
]

\node[db_shape] (AAS) at (8, 1) {AAS};

\node[layer_box, minimum height=5cm] (ContextLayer) at (0,0) {};
\node[layer_box] (BehaviorLayer) at (0,-5) {};
\node[layer_box, minimum height=3.5cm] (CommLayer) at (0,-7) {};

\node[label_style, left=0.5cm of ContextLayer.west] {Context};
\node[label_style, left=0.5cm of BehaviorLayer.west] {Behavior};
\node[label_style, left=0.5cm of CommLayer.west] {Communication \\Manager};

\node[module] (AAShandler) at (6, -1) {AAS capability description handler};
\node[module] (dots0) at (14, -1) {\dots};

\node[module] (Optimizer) at (2,-2.5) {Optimizer};
\node[module] (Constraint) at (6,-2.5) {Constraint-Check};
\node[module] (Execution) at (10,-2.5) {OPC UA Execution};
\node[module] (dot1) at (14, -2.5) {\dots};

\node[module] (MultiRequester) at (2,-4) {Multi-stage Requester};
\node[module] (SingleStage) at (6,-4) {Singlestage Requester};
\node[module] (MultiProvider) at (10,-4) {Multi-service provider};
\node[module] (dot2) at (14, -4) {\dots};

\node[module] (ServiceRequester) at (2,-6) {Service-Requester};
\node[module] (ServiceProvider) at (6,-6) {Service-Provider};
\node[module] (Proxy) at (10,-6) {Proxy};
\node[module] (dot3) at (14,-6) {\dots};

\node[module] (Bidding) at (6,-8) {Bidding Protocol};
\node[module] (Dots) at (14,-8) {\dots};

\node[module] (MQTT) at (2,-9.5) {MQTT};
\node[module] (REST) at (6,-9.5) {REST};
\node[module] (XMPP) at (10,-9.5) {XMPP};
\node[module] (Dots2) at (14,-9.5) {\dots};

\draw[arrow] (AAS.south) -- (AAShandler.north);
\draw[mydash] (AAS.south) -- (dots0.north);

\draw[arrow] (AAShandler.south) -- (Optimizer.north);
\draw[arrow] (AAShandler.south) -- (Constraint.north);
\draw[arrow] (AAShandler.south) -- (Execution.north);
\draw[mydash] (AAShandler.south) -- (dot1.north);
\draw[mydash] (dots0.south) -- (dot1.north);
\draw[mydash] (dots0.south) -- (Execution.north);

\draw[arrow] (Optimizer.south) -- (MultiRequester.north);
\draw[arrow] (Optimizer.south) -- (SingleStage.north);
\draw[arrow] (Optimizer.south) -- (MultiProvider.north);
\draw[arrow] (Constraint.south) -- (SingleStage.north);
\draw[arrow] (Constraint.south) -- (MultiProvider.north);
\draw[arrow] (Execution.south) -- (MultiProvider.north);
\draw[mydash] (dot1.south) -- (dot2.north);
\draw[mydash] (dot1.south) -- (MultiProvider.north);

\draw[arrow] (MultiRequester.south) -- (ServiceRequester.north);
\draw[arrow] (SingleStage.south) -- (ServiceRequester.north);
\draw[arrow] (MultiProvider.south) -- (ServiceProvider.north);
\draw[arrow] (MultiProvider.south) -- (ServiceProvider.north);
\draw[mydash] (dot2.south) -- (Proxy.north);
\draw[mydash] (dot2.south) -- (dot3.north);

\draw[arrow] (ServiceRequester.south) -- (Bidding.north);
\draw[arrow] (ServiceProvider.south) -- (Bidding.north);
\draw[arrow] (Proxy.south) -- (Bidding.north);
\draw[mydash] (dot3.south) -- (Dots.north);
\draw[arrow] (Bidding.south) -- (REST.north);
\draw[arrow] (Bidding.south) -- (MQTT.north);
\draw[arrow] (Bidding.south) -- (XMPP.north);
\draw[mydash] (Dots.south) -- (XMPP.north);
\draw[mydash] (Dots.south) -- (Dots2.north);

\draw[arrow] (MQTT.south) -- (8, -11); 
\draw[arrow] (REST.south) -- (8, -11); 
\draw[arrow] (XMPP.south) -- (8, -11); 
\draw[mydash] (Dots2.south) -- (8, -11); 
\node[interface] (I40Int) at (8, -11.3) {};
\node[below=0.1cm of I40Int, font=\bfseries] {I4.0L};

\end{tikzpicture}
}
    \caption{Layered approach of modules for the active AAS part}
    \label{fig:layer}
    \deleted{\vspace{-15pt}}
\end{figure}
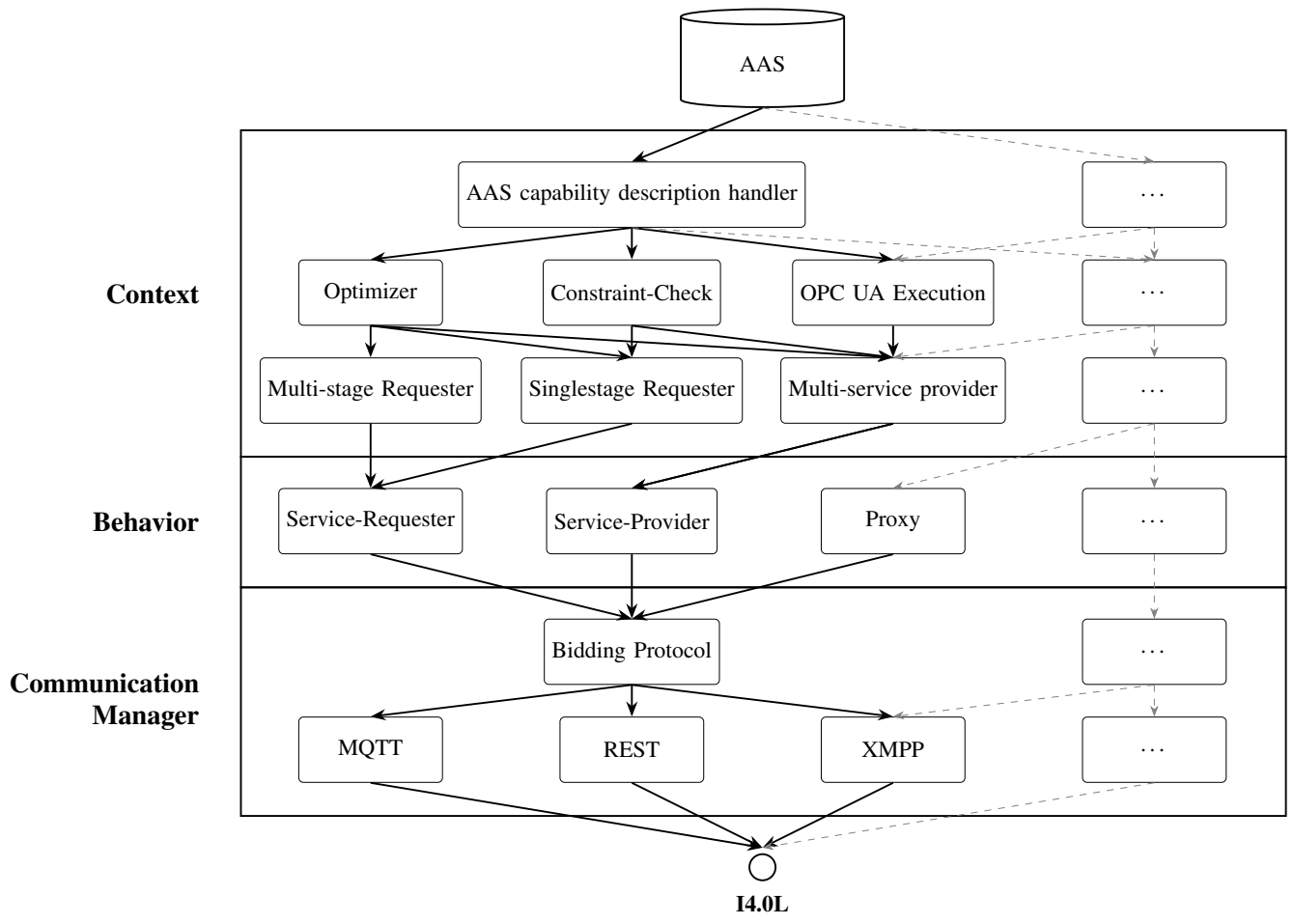

\paragraph{Creating Agent Classes}
\replaced{Upon completion of the analysis phase, the agent classes are created as a first phase based on c). Separation of concerns of the roles and tasks is conducted in the first step, as this is what enables the modularity and reusability required to reuse the same agent classes across heterogeneous assets and network configurations, rather than designing a bespoke agent per use case. Starting with the external view, the messenger (R1) with possible technical interfaces and the interaction manager (R2) assuring conformity to VDI/VDE 2193 protocols are grouped as \textit{CommunicationManager} (CM), since both roles concern how an agent exchanges messages independently of its business logic. The roles related to business behavior, namely the negotiator (R7) for service request and service provision, and the proxy (R4), are grouped as \textit{Behavior}, as they determine what an agent decides rather than how it communicates. The six remaining asset specific roles are grouped as \textit{Context}, since they encapsulate asset dependent information that must be exchangeable without affecting communication or decision logic. This separation is what allows, for instance, a communication manager to be replaced (e.g., a different protocol or technology) without altering the negotiation behavior itself, or the same behavior to be reused across different assets. Possible modules to build up the agent class in the layered approach based on the defined roles are visible in \autoref{fig:layer}. Although the behavior and the communication manager have a more direct connection, the context could be implemented as a separate agent class, which is a precondition for reusing behavior across otherwise heterogeneous assets. The layered architecture can be used to create an agent, such as an MQTT multi-stage service requester, based on the capability description submodel, as part of the system of proactive AAS.}{Upon completion of the analysis phase, the agent classes are created as a first phase based on c).
Separation of concerns of the roles and tasks is conducted in the first step to achieve high modularity and reusability of software components.
Starting with the external view, the messenger (R1) with possible technical interfaces and the interaction manager (R2) assuring the conformity to VDI/VDE 2193 protocols is grouped as \textit{CommunicationManager} (CM). 
The roles related to business behavior, namely the negotiator (R7) for service request and service provision, and the proxy (R4) are grouped as \textit{Behavior}.
The six remaining asset specific roles are grouped as \textit{Context}.
Possible modules to build up the agent class in the layered approach based on defined roles are visible in \autoref{fig:layer}.
Although the behavior and the communication manager have a more direct connection, the context could be implemented as separated agent classes.
The layered architecture can be used to create an agent, such as an MQTT multi-stage service requester, based on the capability description submodel, as part of the system of proactive AAS.}

\paragraph{Constructing Conversations}
\replaced{In the second step of the design process, the conversations are constructed, since a layered agent class only realizes the goals of Section III.a if the layers can actually interact according to the negotiation defined there. The structure of the conversations between the layers for a requester and a provider instance is already shown in \autoref{fig:req} and \autoref{fig:prov}. In terms of design, the conversation structure between them is based on the I4.0L, which is what allows the same conversation pattern to be reused across different implementations of communication manager and behavior. Regarding the conversation flow of a proxy, two communication managers are connected to the behavior, which is what allows different zones or technologies to be bridged without extending the behavior itself. In the simplest case, messages are forwarded with adaptation of the message frame; further extensions may include optimizations such as multi-asset mapping, which require context specific algorithms.}{In the second step of the design process, the conversations are constructed.
The structure of the conversations between the layers for a requester and a provider instance is already shown in \autoref{fig:req} and \autoref{fig:prov}.
In terms of design, the conversation structure between them is based on the I4.0L.
Regarding the conversation flow of a proxy, two communication managers are connected to the behavior. Like that different zones or technologies may be connected.
In the most simple behavior, messages may be forwarded with adaptation of the message frame.
Further extensions may include optimizations like multi-asset mapping requiring context specific algorithms.}

\paragraph{Assembling Agent Classes}
\replaced{In the third design step, the assembly of agent classes and the decentral system is created. Presuming a common domain and semantics of the agents, primarily described by their behavior, a scalable system of proactive AAS can be built, as evaluated in \cite{Dietrich.2026}, because the behavior layer is what different agents must share in order to interoperate, while communication and context can remain heterogeneous. The further integration of the proxy agent class allows requesters and providers to be structured into zones, which is the mechanism by which the number of participating companies in the decentral value network can scale without requiring changes to individual agents. The number of zones, as well as the number of requesters and providers per zone, can therefore be scaled arbitrarily, or in other words implemented independently for every company in the network.}{In the third design step, the assembly of agent classes and the decentral system is created.
Presuming a common domain and semantics of the agents, primary described by their behavior, a scalable system of proactive AAS can be built as evaluated in \cite{Dietrich.2026}.
The further integration of the proxy agent class allows the further structuring of requesters and providers into zones.
The number of zones depending on the number of proxy AAS, as well as the number of requesters and providers per zone, may be arbitrary scaled, or in other words implemented for every company in the decentral value network.}

\paragraph{Deployment Diagrams} 
\replaced{As the final step of the design and of the entire MaSE phases, deployment diagrams are created, since the layered separation only provides practical benefit if it can be mapped onto different deployment environments without redesign. Referring to the architectural realizations introduced in \cite{Stolze.2024b}, all approaches, namely monolithic, agentic, and orchestration, can be realized based on the layered approach precisely because communication manager, behavior, and context are separated: this is what allows the respective software components to be integrated in an AAS server, an AAS repository, as a separate application, or in a standalone AAS itself, depending on which deployment a given company can support. Integration into a standalone AAS itself, e.g., via a container as proposed in \cite{Ellwein.2025}, is particularly relevant, as it could pave the way to self-planning products. Depending on the configuration, MQTT brokers may be required for respective zones, or passive AAS based on, e.g., BaSyx, may be required in the deployment.}{As the final step of the design and the entire MaSE phases, deployment diagrams are created.
Referring to the architectural realizations introduced in \cite{Stolze.2024b}, all approaches, namely monolithic, agentic, and orchestration, can be realized based on the layered approach.
Due to the separation of the communication manager, the behavior, and the context, respective software components may be integrated in an AAS server, an AAS repository, as a separate application, or in a standalone AAS itself.
Especially integration into a standalone AAS itself, e.g., via a container as proposed in \cite{Ellwein.2025}, could pave the way to self-planning products.
Depending on the configurations, MQTT-brokers could be required for respective zones or passive AAS based on, e.g., BaSyx may be required in the deployment.}
















\section{Realization} \label{sec:real}

\begin{figure}
    \centering
    \includegraphics[width=0.7\linewidth]{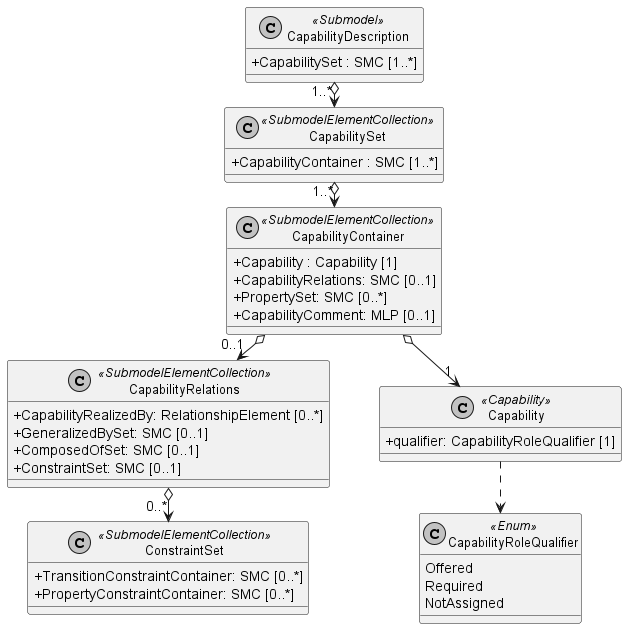}
    \caption{Excerpt of IDTA02020 - Capability Description \cite{idta02020capabilities}}
    \label{fig:cd}
    \deleted{\vspace{-15pt}}
\end{figure}

\replaced{Following}{After} the methodological development of the proactive AAS, this section \replaced{details its}{follows the} realization\replaced{. It}{ and} is structured according to \autoref{fig:layer}\added{, progressing} from the standardized AAS to \added{the }I4.0L.
\paragraph{AAS\replaced{ M}{-m}odeling}
The capability description submodel (cf. \autoref{fig:cd}) \added{detailed }in \cite{idta02020capabilities} serves as \replaced{the}{a} foundation for service modeling.
The capabilities of an asset are organized into sets \replaced{based on}{for} different use cases, each \replaced{featuring}{with} containers that hold individual capabilities.
Each container describes an offered or required capability using a common identifier\replaced{. It}{ and} also includes relationships between capabilities, a set of arbitrary properties, and a human-readable description.
Properties are\added{ either} contained in the submodel itself or \deleted{may }refer to identical properties, such as those defined in the technical data.
The relationships of a capability can include details about its realization, \replaced{such as}{for example} links to the skills of a control component\replaced{. They can also include}{, as well as} references to other capabilities that either generalize it or are used to compose it.
Finally, the relationships of the capabilities include constraints on properties or transitions, which may be defined as pre-\added{conditions}, invariant\added{s}, or post\added{-conditions}.

\replaced{As identified, services primarily extend capabilities in commercial aspects. Therefore, encoding the service information model in \cite{Dietrich.2026} into the standardized submodel is more suitable for achieving a minimum viable model, than creating a custom, non-standardized submodel.}{ 
To achieve a minimum viable model of services in the standardized submodel, the information of \cite{Dietrich.2026} is integrated and extended as follows.}
The semantic description of the service $\sigma$ is integrated in the \textit{idShort} of the \textit{Capability} element.
The provision or request of multiple services is indicated with one \textit{CapabilityContainer} each.
\replaced{To distinguish public and private properties, such as cost, quality or duration, two \textit{PropertySets} are used and indicated by an application specific valuequalifier of type ``private''}{
Properties such as cost, quality or duration are separated into public and private \textit{PropertySets}, indicated by a custom valuequalifier of type ``private''.}
This allows the separation of public properties to be sent as interaction elements in the \textit{ServiceRequest} of a requested capability (cf. \autoref{fig:req} No. 8-12) or the \textit{ServiceOffer} of a provided capability (cf. \autoref{fig:prov} No. 12-16) and private properties referenced in a \textit{PropertyConstraintContainer}.
As an example, this also allows for a distinction between private manufacturing costs and public distribution costs for a provider.
Boundaryconditions, e.g., to realize the capability based on quality or the feasibility based on costs, are integrated into the basic property constraint as a precondition.
Out of possible \textit{PropertyConstraintType} a \textit{BasicConstraint} is chosen to model the boolean logic.
It includes the value type to be compared and a qualifier of the boolean logic as provided by the AASX-Package Explorer.
Furthermore, a relationship indicates the connection between the constraint type and the properties already addressed.
To enable the multi-stage approach, the \textit{TransitionConstraintContainer} is used.
It is indicated as a precondition in the \textit{TransitionConditionalType} and contains a relationship \replaced{between}{of} the capability itself to the constraining capability.
To determine whether a prerequisite capability has been met, the boolean value of the \textit{ValueQualifier} in the \textit{Capability} of type ``required'' is set to ``false'' after execution.
In the register of an AAS\added{,} a multi-stage service request is validated by extracting capabilities and their preconditions into a graph\added{,} followed by the analysis of the graph's topology (cf. \autoref{fig:req} Nos. 1-4).
\replaced{C}{The c}ompliance \replaced{with}{to} the metamodel itself\added{,} as well as the validation\added{,} is carried out by the AAS capability description handler based on\added{ the} \textit{semanticID} of the submodel.

\paragraph{Constraint Check} 
The approach of the constraint check based on the \textit{BasicConstraints} presumes a common semantics on the requester and provider side\added{s}.
When receiving a service request, the provider identifies its capable services based on their constraints (cf. \autoref{fig:prov} No\added{s}. 4-7), similar to the requester analyzing the feasibility of an offer based on the constraints of the required service (cf. \autoref{fig:req} Nos. 15-18).
In these steps, the module performing the constraint check queries the interaction elements of an incoming I4.0L message for properties with\added{ an} identical \textit{idShort}.
If one property is present, the tuple of\added{ the} internal constraint value, constraint type including\added{ a} boolean operator, and queried property is built.
For numeric values, the resulting equation is evaluated.
In\added{ the} case of string values, only an equal\added{ity} operation or \deleted{a }containment in a list based on the ``FormChoices'' qualifier is supported.

\paragraph{Optimizer}
After detailing the constraints, an optimum of one or multiple requests to one or multiple offerings needs to be identified, likewise on\added{ the} provider (cf. \autoref{fig:prov} No\added{s}. 8-9) and requester (cf. \autoref{fig:req} No\added{s}. 19-20) side\added{s}.
In contrast to a sole first-in first-out (FIFO) processing of requests and offers, this enables the integration of optimization mechanisms in\added{to} the behavior.
This work introduces two algorithms for the best provider fit and a requester lowest-cost optimization, both depending on the expected response time of the I4.0L message as a deadline \replaced{for}{of} the decided response.

The best\replaced{-}{ }fit optimization focuses on the provider\added{'s} decision to offer capable services within the workload limit to one or multiple incoming requests within their response time.
At that time, the requester's budget is expected to be unknown, which only allows a decision \added{based }on the organizational or potential technical properties of the request, summarized as the quality of the service vector $\mathbf{q}_j^{req}$.
Properties with the same \textit{idShort} or \textit{semanticID} are expected in the provided services, \replaced{represented}{resembled} as $\mathbf{q}_i^{prov}$.
For each combination $(i,j)$, the distance of \added{the }quality of service $\lvert\mathbf{q}_j^{req}-\mathbf{q}_i^{prov}\rvert$ is calculated.
In ascending order, services are provided to requests within the possible workload.
This optimization assumes that a best\replaced{-}{ }fit quality provides an attractive solution instead of an exceeding one.

The lowest\replaced{-}{ }cost optimization focuses on the requester\added{'s} decision to choose one of the multiple offers for a specific request, based on the cost factors provided $\mathbf{c}_i$.
Based on a private weighting vector $\mathbf{w_j}$ in the properties of the requested service, the optimal offer $i$ for a request $j$ is chosen based on a weighted minimum $\min(\mathbf{w_j}\cdot\mathbf{c}_i)$.

\paragraph{Multi-stage requester and multi-service provider}
The AAS modeling and the optimization possibilities are combined in the context of\added{ a} multi-stage requester and\added{ a} multi-service provider.
Based on a common context class, they allow the arbitrary integration of constraint, optimization, or execution modules and connect single- or multi-stage requests to the service negotiation.
The multi-stage requester includes the respective modules and adds \deleted{the }further logic to determine services \replaced{for}{of} which all \textit{TransitionConstraintContainers}\added{,} including preconditions\added{,} are met based on the mentioned qualifier.
In addition, once the service is complete\added{d}, it adjusts the qualifier and calculates the total cost of a multi-stage service—for example, to determine the total cost of a product.
The multi-service provider, on the other hand, integrates\added{ the} respective modules and is responsible for \deleted{the }workload management.
This means determining whether a capable service is actionable based on a private property indicating its capacity.
Only if the number of reserved services (see \autoref{fig:prov} Nos. 8, 25, 34) in the context is below capacity, the service is considered actionable.

\paragraph{Provider, requester and proxy behavior}
The negotiations of the single services derived from the context are carried out by the behavior.
Regarding\added{ the} proxy only, message forwarding, as explained in \autoref{sec:meth}, is implemented in a possible configuration.
On the one hand, the behavior of the requester is responsible for creating requests with unique conversation \replaced{IDs}{ids}, collecting offers, trigger\replaced{ing}{s} their evaluation within response times to create orders or reject\replaced{ions}{s}, and process\replaced{ing}{es} the result\added{s}.
Its configurations are the expected response time, the limit of retries in \added{the }case of no feasible offers, and a retry wait time until the next \textit{ServiceRequest} is sent.
On the other hand, the behavior of the provider collects incoming requests, triggers their evaluation within\added{ the} response time to create offers or refus\replaced{als}{es}, create\added{s} service instances for execution, and provide\added{s} execution results.
The behavior can be configured with the expected response time and the type of offer generation, e.g.\added{,} if \deleted{a }FIFO processing or a best fit should be used.

\begin{figure}
    \centering
    \includegraphics[width=\linewidth]{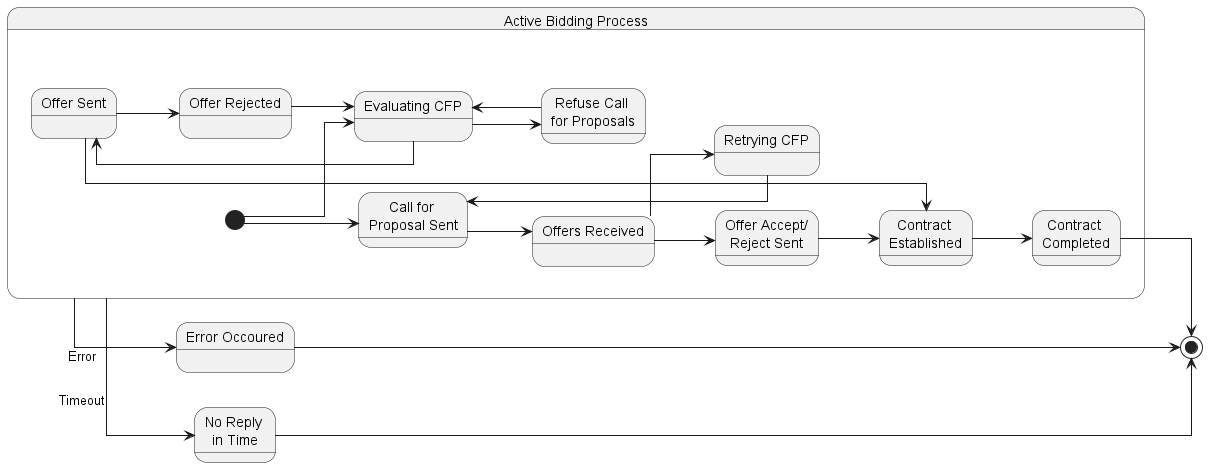}
    \caption{State chart of the bidding protocol}
    \label{fig:states}
    \deleted{\vspace{-15pt}}
\end{figure}

\paragraph{Bidding protocol}
The underlying negotiation protocol is defined in the bidding protocol according to \cite{VereinDeutscherIngenieure.c}.
\added{It }integrates the messages of the I4.0L and tracks negotiations based on a state chart derived from the protocol (cf. \autoref{fig:states}).
The state chart is applicable \replaced{to}{for} both\added{ the} requester and\added{ the} provider, and allows their behavior to update the states and check whether incoming messages are provided within response times.

\paragraph{Messenger}
The last element is the messenger.
Implementations for specific technologies are based on a parent class integrating \deleted{the }abstract functions to (i) send I4.0L\replaced{-compliant}{ conform} messages as\added{ a} broadcast or to chosen receivers, (ii) receive and validate messages and forward them to\added{ the} respective handle\added{r}s of their type, (iii) track metrics like the number of incoming or outgoing messages, and (iv) provide a unique identification.
Based on this, an MQTT messenger is realized.
The I4.0L \deleted{compliant} message is sent as\added{ the} payload of an MQTT message, whilst the topic of the message is differentiated between\added{ a} broadcast for, e.g., \textit{ServiceRequests} and the \replaced{ID}{id} of the messenger for direct messages in the following negotiation states.

\replaced{The proa}{A}ctive AAS is implemented as a modular \replaced{P}{p}ython package based on aas-core-works\footnote{https://github.com/aas-core-works/}. 
With \replaced{this}{the} modular and package\replaced{-}{ }based approach, it supports \deleted{the }reusability and may be integrated in\added{to} AAS\replaced{ }{-}servers, custom orchestration software or containerized standalone AAS.

\section{Exemplification} \label{sec:exe}

\begin{figure}[t]
    \centering
    \includegraphics[width=\linewidth]{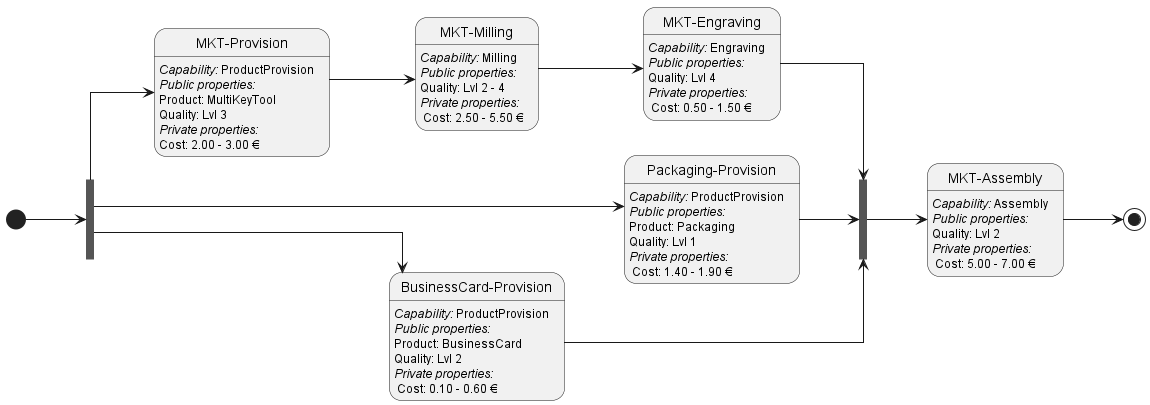}
    \caption{Priority graph of the products multi-stage service request}
    \label{fig:mkt}
\end{figure}

Based on the realization presented, a functional prototype of the \added{pro}active AAS is implemented for the manufacturing-x testbed at ISW presented in \cite{Clar.2026}.
The purpose of the testbed is to individually manufacture configurable products of Multi-Key-Tools (MKT) with several production, assembly and logistic steps within a flexible value network.
The testbed is built on multiple demonstrators each considered as individual compan\replaced{ies}{y} providing relevant \deleted{relevant }services to build up the value stream.
For this exemplification, the central manufacturing and assembly steps of the individual products are considered as required services.
The priority graph of the services required that includes the relevant information is shown in \autoref{fig:mkt}.
Cost and quality level factors are scaled based on user-defined parameters within the intervals, with levels one to five considered for quality.


The companies considered offering services are the warehouse that provides the different parts, a manufacturer, an engraving, and an assembly-company.
Instead of an OPC UA connection on the control components, the \replaced{skill}{process} execution is abstracted by its duration.
The relevant properties of their services are shown in \autoref{tab:prov}.
The costs are based on purchase prices and estimates, whilst quality levels are assumed.
The total capacity of the warehouse is limited to three simultaneous services in the context configuration.
\begin{table}[t]
    \centering
    \caption{Overview of service providing companies}
    \resizebox{\linewidth}{!}{
    \begin{tabular}{p{1.2cm}  p{1.1cm} | p{1.45cm} | p{1.45cm} | p{1.45cm} | p{1.45cm} | p{1.45cm} | p{1.45cm}}
        Category & Parameter & \multicolumn{3}{|c|}{Warehouse} &  Manufacturer & Engraver & Assembler\\
        \midrule
        \multicolumn{2}{l|}{Capability} & \multicolumn{3}{|c|}{ProductProvision} & Milling & Engraving & Assembly\\ 
        \midrule
        Public properties & Cost & 2.38 € & 1.56 € & 0.20 € & 3.75 € & 0.9 € & 5.8 € \\
        \midrule
        Private & Quality & Lvl 3 & Lvl 2 & Lvl 3 & Lvl 3 & Lvl 5 & Lvl 3  \\
        properties & Duration & 45 s & 45 s & 45 s & 80 s & 20 s & 170 s \\
        & Product & MultiKeyTool & Packaging & BusinessCard & - & - & - \\
        & Capacity & 2 & 1 & 3 & 2 & 3 & 4 \\
        \midrule
        \multicolumn{2}{l|}{Conditions} & \multicolumn{3}{|c|}{Product, Quality} & Quality & Quality & Quality
    \end{tabular}}
    \label{tab:prov}
    \deleted{\vspace{-15pt}}
\end{table}

The total duration of the critical path sums up to $315$ seconds based on provider properties.
The corresponding product AAS containing the multi-stage service requests and the companies AAS including the offered services according to the standardized submodel are available on \replaced{GitHub}{github}\footnote{Examples available at: \url{https://github.com/ARENA2036-Well-defined/aas-examples-service-negotiation}}.

To demonstrate the developments and analyze the influence of the \added{proposed} proxy, a deployment is set up on a virtual machine in Proxmox with Ubuntu 24.04, 90GB RAM, \added{and }32 x86-64-v4 processors.
It contains all the AAS in one BaSyx portfolio (environment, registry and discovery), a Mosquitto MQTT broker and active AAS applications for each of the four provider AAS, one for the product requests and one as proxy, all deployed on \replaced{D}{d}ocker.
For the four providers, a response time of $10s$ is set, the requester configuration has a response time of $5s$, maximum 30 retries, and a retry wait time of $15s$.
For five different product AAS, \replaced{including}{containing} one \replaced{i}{u}ncapable and one unfeasible solution, the following configurations are performed.
In Configuration I, all messengers are within one zone and providers create offers based on the FIFO principle.
In configuration II one proxy forwards between the requester and a zone of all providers, and in configuration III the best-fit provision is activated without the usage of zones.
The resulting time to complete all multi-stage service requests, leading to finished or \replaced{incomplete}{uncompleteable} products, \replaced{was}{were} verified in multiple runs and \replaced{is}{are} displayed in \autoref{tab:result}.
In each case all possible services were executed.

\begin{table}[t]
    \centering
    \caption{Resulting simulation time for the three configurations}
    \begin{tabular}{l c c c}
        Configuration & I & II & III \\
        \midrule
        Total simulation time & 547.8 s & 548.0 s & 487.7 s 
    \end{tabular}
    \label{tab:result}
    \deleted{\vspace{-15pt}}
\end{table}

\added{The results show the feasibility of the realized proactive AAS.}
A comparison of the results of configuration I and II shows that the additional use of the proxy has no significant effect on simulation time.
It thus enables scalable systems by bridging different zones or protocols and the possible integration of further access logic or management capabilities.
Comparing the results of configuration I and III, the presented best-fit provider behavior shows a significant impact compared to FIFO principles, when providers deal with multiple requests at once.
In the example, the better fitting quality of service for the MKT-provision, e.g., resembling delivery time, leads to an optimized scheduling of the critical path in the decentralized system.
This aspect needs further consideration in the future to realize scalable decentral\added{ized} systems.
For both, the modular, layer-based approach provides a targeted foundation.




\section{Conclusion and Outlook} \label{sec:conc}


This work introduced a methodologically grounded, MaSE-based design for \added{pro}active AAS\replaced{. This design}{ that} separates communication, negotiation behavior, and asset-specific context, thereby providing a reusable pattern for decentralized service negotiation in heterogeneous value networks.
Building on this design, we realized a capability modeling approach \replaced{using}{with} the IDTA 02020 submodel\replaced{. This approach}{ that} embeds multi-stage dependencies, public/private properties, and constraint checks\added{ directly} into the AAS itself\replaced{Furthermore, we}{ and} combined it with configurable negotiation and optimization strategies on requester and provider side\added{s}.
\replaced{Compliance with}{The compliance to} the I4.0L specification is achieved \replaced{through}{by the use of} standardized AAS submodels \replaced{and a}{as well as the} communication manager \replaced{which ensures correct message structure}{ensuring the structure of messages} for various technical implementations and \deleted{the used} interaction protocol\added{s}.
\replaced{The evaluation of the functional prototype in the Manufacturing-X testbed demonstrated the decentralized planning of cross\replaced{-}{ }company value chains \replaced{using}{with} proactive AAS. Furthermore it shows, the tangible impact on completion time of the negotiation behaviors as part of the proactive AAS design. Last the compliant use of standards for\added{ both} the capability description submodel and the I4.0L was confirmed.}{The exemplary evaluation in a Manufacturing-X testbed demonstrated that these design and modeling choices are implementable with existing AAS and I4.0L standards and that different behaviors have a tangible impact on completion time and scalability.}
\added{In comparison to the state of the art in \autoref{tab:sota}, the presented results meet\deleted{s} all requirements except for the scalability evaluation.}

However, the application of MaSE in this work is limited\replaced{.}{:} \replaced{N}{n}ot all MaSE phases are exploited end-to-end, \added{and }implementation details and algorithms are only partially traced back to MaSE artefacts\replaced{. Additionally,}{, and} goals, roles, and conversations are simplified, focusing on the bidding protocol \replaced{while omitting}{and omitting} many exceptional and non-functional aspects.
The agent decomposition was validated only in one testbed with a specific MQTT- and proxy-based topology, and alternative decompositions or zoning concepts were not systematically compared.
Furthermore, MaSE was applied mainly from a technical perspective, so business and semantic requirements (e.g., contractual variants, rich cost models) remain only coarsely represented.
In addition, the approach assumes shared semantics across companies, uses simplified cost and quality models, and employs basic optimization heuristics with fixed response times.
Future work should therefore extend business and cost modeling, strengthen semantic interoperability via shared vocabularies and ontologies and systematically explore alternative MaSE-based decompositions.
Furthermore \replaced{future research could integrate}{an integration of} more advanced scheduling and negotiation mechanisms, such as auction-based and online optimization methods, to support larger and more heterogeneous industrial ecosystems\replaced{. In addition,}{ could be performed and} other standardized constraint types and submodels, like referred technical data, could be included in the active AAS.
\added{Last, a systematic scalability evaluation similar to the preliminary analysis in \cite{Dietrich.2026} has to be conducted for the resulting system.}


\section*{Acknowledgment}

This research and development project is funded by the German Federal Ministry of Research, Technology and Space (BMFTR) within the ``Research Campus – Public-Private Partnership for Innovation'' funding initiative (02P23Q820 ff) and managed by the Project Management Agency Karlsruhe (PTKA). The authors are responsible for the content of this publication.


\bibliographystyle{IEEEtran}
\bibliography{bibliography}

\end{document}